\documentclass[floatfix,11pt,onecolumn,showpacs,amsmath,amssymb]{revtex4}

\usepackage{epsf}
\usepackage{graphicx}  
\usepackage{dcolumn}   
\usepackage{bm}        

\newcommand{\be}{\begin{equation}}
\newcommand{\en}{\end{equation}}
 \newcommand{\bea}{\begin{eqnarray}}
 \newcommand{\ena}{\end{eqnarray}}
  \newcommand{\sch}{Schwarzschild}

\begin{document}

\title{A New Property of the Electromagnetic/Yang-Mills-Conformal Gravity System  in Spherical Symmetry}
\author{Hongsheng Zhang$^{1,2~}$\footnote{Electronic address: sps\_zhanghs@ujn.edu.cn} }
\affiliation{$^1$ School of Physics and Technology, University of Jinan, 336 West Road of Nan Xinzhuang, Jinan, Shandong 250022, China\\
$^2$ State Key Laboratory of Theoretical Physics, Institute of Theoretical Physics, Chinese Academy of Sciences, Beijing, 100190, China}

\date{ \today}

\begin{abstract}
 We find a new property in $W^2$-conformal gravity in spherical symmetry. We demonstrate that the charge of the electromagnetic field varies with respect to the partial scaling symmetry (conformal transformations in subspaces of a spacetime) in the conformal gravity. We find that the electric and magnetic charges can vanish or appear via  partial conformal transformations for the on-shell
 configurations in the conformal gravity. We find a solution of a Yang-Mills SU(2)-charged black hole in the conformal gravity, and further demonstrate that the SU(2) charge has similar properties with electromagnetic magnetic charges  in the conformal gravity. We demonstrate how to imbed the electromagnetic and Yang-Mills SU(2) gauge field into the conformal gravity via such a symmetry. We make a brief discussion about the possibility to extend this argument to the Einstein gravity.

 \end{abstract}


\preprint{arXiv: }
 \maketitle

    In the late 1910's, Weyl claimed that the gravity and electromagnetism interactions could be unified by an ``infinitesimal geometry" \cite{weyl}. In modern language, Weyl's infinitesimal geometry is defined as a pair of a metric $g$ and an 1-form $\lambda$ on a smooth manifold $M$,
    \be
    W=(g_{ab}, \lambda_{c}),
    \en
    with the (Weyl)-derivative operator,
    \be
    \nabla_{c} g_{ab}=-2\lambda_{c}g_{ab},
    \label{wey}
    \en
    where $W$ is also called the Weyl structure.
    It is easy to show that (\ref{wey}) is invariant under the Weyl transformations,
    \be
    \bar{g}_{ab}=e^{2\xi}g_{ab},~~~\bar{\lambda}_c=\lambda_c-(d\xi)_c,
    \label{tra}
    \en
    where $\xi$ is a smooth function on $M$. Further, one defines an electromagnetic field by using the 1-form $\lambda$,
    \be
    {\cal F}=d\lambda.
    \label{ele}
    \en
    Thus a Weyl structure corresponds to a self-consistent system.  In such a system for the self-consistency a conformal
    transformation imposed upon the metric $g_{ab}$ requires a gauge transformation for the electromagnetic field $F$. We emphasize that
    the Weyl transformations (\ref{tra}) can not change an electromagnetic field. In fact the electromagnetic
    field $\cal F$ is invariant under such a transformation, while the gravity field $g_{ab}$ gets ``essentially" (in view of the Einstein gravity) modified. The whole effect is that the
    Weyl structure remains unchanged, that is, the equations (\ref{wey}) and (\ref{ele}) hold after the transformation (\ref{tra}). In this paper we will demonstrate that conformal transformations in the subspaces of a spacetime with {\emph {constant scale factors}} really alter electromagnetic fields.

    The Weyl structure (\ref{wey}) clearly displays that the derivative operator is not compatible with the metric. The result is
    that the length of a parallelly transported vector will change.  Two same clocks put at the same point A initially will run at different rates (not the synchronization problem) when they reach the point B, if they travel along different paths. This point  leads to a conclusion that  hydrogen atoms which are separated and then put together will have different spectra in a magnetic field, for example the geomagnetic field. But we never observe such a
    discrepancy. This is Einstein's objection to the Weyl theory. In this paper, we will always adopt metric-compatible derivative operators. And thus Einstein's argument is untenable in our scenario.

   Though Weyl's original effort to unify gravity and electromagnetic fields fails, it gets great success when one replaces the conformal transformation
   of the metric by a rotation in the intrinsic space of a matter field, which is the gauge theory.  In this approach the transformations are done in the intrinsic space rather than in spacetime, and the implemented subject is the wave function of a matter field rather than the spacetime metric. And thus London can show that electromagnetic is a U(1) gauge field accompanied with phase transformations a matter field, especially a fermionic field \cite{london}, see also in a review \cite{review}. Then Weyl systematically investigated the phase transformation and showed that electromagnetic field is a gauge field \cite{weyl2, review}. Yang and Mills extended the U(1) gauge theory to a non-abelian one \cite{yangmills, review}.  The non-abelian Gauge theory is the foundation of the electro-weak interaction and the strong interaction (the standard model of particle physics). Also, the original idea of  conformal transformations of the metric has been revisited several times in
   different contexts of theoretical physics \cite{olz}.

    Conformal gravity is a prominent one in several aspects in modified gravities. At classical level all the vacuum solutions of Einstein gravity are solutions of the conformal gravity. In this sense the observational verifications of general relativity can be treated as verifications of conformal gravity. There exist some configurations which are solutions of conformal gravity but not the Einstein gravity, which makes special predictions of conformal gravity. For example, in the spherically symmetric case, a solution has been found in \cite{bir}, which is an extension of the \sch~solution. This solution is demonstrated to be helpful to interpret the   surplus rotation velocity of the celestial objects far from the center of galaxies \cite{man1, man2}, i.e., the dark
   matter problem in general relativity. Further, it is demonstrated that one can pick up the solutions of the Einstein gravity from all the solutions of conformal gravity by imposing some special boundary conditions \cite{mal}. At quantum level, the conformal gravity also has some significant properties deserving to study further. It is wellknown that the Einstein gravity is not renormalizable.   Remarkably,  at one loop level, the UV divergence is tractable for conformal gravity in 4-dimensional spacetime. At two-loop level, the UV divergence becomes untractable and the theory is not renormalizable \cite{FTC}. Recently it is shown that the conformal gravity will naturally appear when one requires the property of renormalization of the Einstein gravity \cite{hoo}. The infared problem also very important in modern field theory. The infared problem in conformal gravity may be more deep and complicated. A matter field may induce a linear Einstein term (Ricci scalar) at some low energy region \cite{LE}. The problem of unitarity is a general problem in higher order derivative theory. It is conjectured that the theory of conformal gravity becomes very different from the linearized theory, in which the case is something like the quark confinement in QCD, and the unitarity may be recovered  at the non-perturbative level \cite{UNITR}. Some exact solutions in the conformal gravity are presented in \cite{liu}. The conformal gravity in the scenario of AdS/CFT  is studied in \cite{gru}. Quantum conformal gravity and singularity problem is investigated in \cite{quantumcon}. Formation and evaporation of charged black holes are explored in \cite{formcon}.

   In modern physics, the conformal symmetry is more and more important.
   In modern (classical) field theory, a massless particle has the symmetry of conformal invariance.
   If the energy scale is high enough, all the fields are conformally invariant in the standard model (quantum anomaly is not included).
   In view of these facts, the conformal field theory is extensively applied in many different areas of physics, including the string theory, statistical physics, and condensed matter physics etc \cite{many}.
    However gravity theory, which is the original theory investigated by using conformal symmetry, makes a dramatical exception. The Einstein gravity is not conformally invariant, and this property is independent on the energy scale.

   Imposing the
   condition of conformal invariance, one finds that the unique gravity is,
    \be
   S=\int d^4x \sqrt{-g}~ \left(-\frac{\alpha}{16\pi} W_{abcd}W^{abcd}+{\cal L}_m\right),
   \label{act}
   \en
   where ${\cal L}_m$ denotes the Lagrangian of the matter fields. We consider a system composed of electromagnetic (U(1))
   and SU(2) gauge fields,
   \be
  {\cal L}_m=-\frac{1}{4}\left({\cal F}_{ab}{\cal F}^{ab}+\frac{1}{g_s^2}F^A_{ab}F^{Aab}\right),
  \label{lag}
  \en
  where ${\cal F}$ denotes the electromagnetic field and $F$ denotes the Yang-Mills field, and a capital Latin letter indicates the group index. $g_s$ denotes the coupling constant of the SU(2) field, while for the
  abelian gauge field we can always rescale the coupling constant to $1$.
    The corresponding field equation reads,
   \be
   -4\alpha C_{ab}=8\pi T_{ab},
   \label{bach}
   \en
   where $T_{ab}$ is the stress-energy corresponding to ${\cal L}_m$, and $C_{ab}$ is the Bach tensor,
   \be
   C_{ab}=(\nabla^c\nabla^d+\frac{1}{2}R^{cd})W_{acbd}.
   \en

    In this paper we will discuss the conformal transformation in a subspace of a manifold. The result shows that the electromagnetic and
    SU(2) Yang-Mills charges
    appear or vanish after such a transformation in the conformal gravity.

    First, we consider the spherically symmetric case.  The charged spherically symmetric solution in the conformal gravity reads \cite{bir},
  \be
  ds^2=-fdt^2+f^{-1}dr^2+r^2d\theta^2+r^2\sin^2\theta d\phi^2,
 \label{schl}
  \en
  where
  \be
  f=\frac{h}{r}+i+jr+kr^2.
  \label{fhi}
  \en
  Here $(t,~r,~\theta,~\phi)$ are (\sch-like) spherical coordinates, and $h,~i,~j,~k$ are four integration constants. Turning off the Yang-Mills field, we write
  electromagnetic potential as,
     \be
   {\cal A}=\frac{q}{r}dt-v\cos\theta d\phi.
   \label{pot}
   \en
   One can check that this potential satisfies the Maxwell equation.
  And from the field equation (\ref{bach}), one finds that the electric charge $q$ and magnetic charge $v$ satisfy,
  \be
  q^2+v^2=\frac{2\alpha(i^2-1-3hj)}{3 }.
  \label{qv}
  \en
  Generally, a conformal transformation is defined as
  \be
  \widetilde{ds}^2=e^{2\epsilon}ds^2.
  \en
  We name it ``total" conformal transformation. If $\epsilon$ is constant, the transformation is a
  global transformation. If $\epsilon$ is a function on the manifold, the transformation is a local one.
  One should not confuse ``total" transformation with ``global" transformation.

   Now we define the
  partial transformation. Besides the total conformal transformation, one can make  conformal transformation  in a submanifold of the spacetime,
  \be
  \widetilde{dX}^2=e^{2p}dX^2,
  \en
  where $dX^2$ is the line element of a submanifold of the spacetime. We name such a transformation partial conformal transformation. The corresponding symmetry
  is named as partial scaling symmetry, or partial conformal symmetry.

  We make partial conformal transformations in the subspaces $dt,~dr,~d\theta,{\rm and}~d\phi$ simultaneously,
  \be
  \widetilde{~ds~}^2=-f\mu_0dt^2+f^{-1}\mu_1dr^2+r^2\mu_2d\theta^2+r^2\sin^2\theta \mu_3d\phi^2,
  \label{tmet}
  \en
  where $\mu_0,~\mu_1,~\mu_2,~\mu_3$ are four constant conformal factors. Hence they are global transformations. We emphasize that they are not coordinate transformations.  Formally this map is,
  \be (M,~g,~\nabla,~{\cal A})\to~ (M,~\tilde{g},~\tilde{\nabla},~{\cal A}).
  \label{trans1}
  \en
   In this map, $M$ is the base manifold in consideration,
  $\tilde{g}$ is presented in (\ref{tmet}), the derivative operators $\nabla$ and $\tilde{\nabla}$ are compatible with the
  metrics $g$ and $\tilde{g}$ respectively, and ${\cal A}$ denotes the electromagnetic potential, which is formally invariant in the transformations. However, one will see that ${\cal A}$ describes different electromagnetic fields before and after transformations.

  By comparison, Weyl's original suggestion is a map,
  \be
  (M,~g,~\nabla,~{\cal A})\to~ (M,~\tilde{g},{\nabla},~\tilde{{\cal A}}).
  \en
   In Weyl's map, the derivative operator $\nabla$ is invariant, while the
   metric undergoes a conformal transformation. Thus they are not compatible before, or after, or both with respect to the conformal transformations. The length
   of parallelly transported vector will depend on the pathes, which is excluded by experiments. This is the essence of Einstein's objection to Weyl's theory, as
   a postscript in Weyl's original paper.

    In our suggestion, we always use the metric-compatible derivative operators before and after transformations. So
   Einstein's objection does not work. The other point deserved to note is that the electromagnetic field is invariant since ${\cal A}$ just gets a
   gauge transformation in Weyl's map. While in our suggestion, the electromagnetic field gets essential change, though ${\cal A}$ is formally unchanged.

   The $W^2$-term is invariant under total conformal transformations originally suggested by Weyl. Correspondingly, we consider the  variation of the action under such partial conformal transformations before investigating the field equation. Before partial conformal transformations, the $W^2$-term yielded by (\ref{schl}) reads,
   \be
   W_{abcd}W^{abcd}=\frac{(r^2f''-2rf'+2f-2)^2}{3r^4}.
   \label{w21}
   \en
   After the partial conformal transformations, the $W^2$-term yielded by (\ref{tmet}) reads,
   \be
    \widetilde{W}_{abcd}\widetilde{W}^{abcd}=\frac{(\mu_2r^2f''-2\mu_2rf'+2\mu_2f-2\mu_1)^2}{3\mu_1^2\mu_2^2r^4}.
    \label{w22}
    \en
  If $\mu_1=\mu_2$, (\ref{w22}) recovers (\ref{w21}) up to a constant factor. In general, they are different. However, with aid of gauge fields, their on-shell configurations possess the following relations.

  Using (\ref{tmet}), it is easy to obtain the stress energy after the transformation,
  \be
   -8\pi \rho=8\pi T_0^0=\frac{2\alpha(\mu_1^2- \mu_2^2 i^2+3 \mu_2^2 h j)}{3 \mu_1^2\mu_2^2 r^4 },
  \en
  \be
   8\pi p_r=8\pi T_1^1=\frac{2\alpha(\mu_1^2- \mu_2^2 i^2+3 \mu_2^2 h j)}{3 \mu_1^2\mu_2^2 r^4 },
  \en
  and
  \be
  p_\theta=p_\phi=T^2_2=T^3_3=-p_r.
  \en
  The stress energy of the electromagnetic field (\ref{pot}) reads,
  \be
  8\pi T(EM)^0_0=-\frac{{\mu_2} {\mu_3} q^2+{\mu_0} {\mu_1} v^2}{{\mu_0} {\mu_1} {\mu_2} {\mu_3} r^4},
  \en
  \be
  T(EM)^0_0=T(EM)^1_1=-T(EM)^2_2=-T(EM)^3_3.
  \en
  The field equation requires,
  \be
  T=T(EM).
  \en
  A naive result of the above equation is,
  \be
  \mu_2\mu_3q^2+\mu_0\mu_1v^2=-\frac{2}{3}\frac{\mu_0\mu_3}{\mu_1\mu_2}\alpha(\mu_1^2- \mu_2^2 i^2+3 \mu_2^2 h j).
  \label{qvd}
  \en

   The subtle point is that $q$ no longer presents the total charge of the whole space, since the metric and derivative operator have been changed and thus the electromagnetic field $\cal F$ is changed. After the partial conformal transformation (\ref{tmet}) the total charge reads,
   \be
   \tilde{q}=\frac{1}{4\pi}\int *\widetilde{\cal F}=\sqrt{\frac{\mu_2\mu_3}{\mu_0\mu_1}}q,
   \label{ecq}
   \en
   while the magnetic charge is invariant,
   \be
   \tilde{v}=\frac{1}{4\pi}\int \widetilde{\cal F}=v.
   \label{mcv}
   \en
  Substituting (\ref{ecq}) and (\ref{mcv}) into (\ref{qvd}), one arrives at,
  \be
    \tilde{q}^2+\tilde{v}^2=-\frac{2\alpha \mu_3(\mu_1^2- \mu_2^2 i^2+3 \mu_2^2 h j)}{3 \mu_1^2\mu_2  }.
    \label{qvt}
    \en
 $\mu_0$ disappears in the above equation, since a time rescaling can completely cancel it. Thus $\mu_0$ is not a true
 conformal degree for metric, which is coincident with one's intuition about the metric (\ref{tmet}). While $\mu_1,~\mu_2,~\mu_3$ are true physical degrees,
 which lead to physical effects. To understand the transformation properties of the electric and magnetic charges, one needs to observe the transformation properties of the metric and the derivative operator. The potential is invariant, and thus the field gets modified since the field depends on the derivative operator, which uniquely determined by the metric. And the metric undergoes partial conformal transformations, which leads to different derivative operators before and after the transformations.  For the magnetic charge, after an integration, the transformations of the metric and the derivative operator contradict each other, and thus one gets the same charge. For the electric charge, the result becomes different since there is a Hodge dual operation before obtaining the electric charge. The dual operation makes the integration shifts from the $t-r$ subspace to $\theta-\phi$ subspace, and thus the transformations of metric and derivative operator do not contradict each other any more, leading to a factor $\sqrt{\frac{\mu_2\mu_3}{\mu_0\mu_1}}$.      For total conformal transformation, that is, $\mu_0=\mu_1=\mu_2=\mu_3$, from (\ref{ecq}, \ref{mcv}, \ref{qvt}) one obtains
  \bea
      \tilde{q}&=&q, \\
      \tilde{v}&=&v,  \\
      \tilde{q}^2+\tilde{v}^2&=&-\frac{2\alpha (1-  i^2+3  h j)}{3  }.
    \label{qvttotal}
    \ena
    Thus the charges are invariants under total conformal transformations.

 Comparing (\ref{qv}) and (\ref{qvt}), one finds that the quadratic sum of the electric charge and magnetic charge gets shifted under partial conformal transformations.
 Especially, for any triple parameter ($i,~h,~j$) in (\ref{qv}), one can find infinite triples ($\mu_1,~\mu_2,~\mu_3$) to make the quadratic sum of the electric charge and magnetic charge vanish. So the electromagnetic and magnetic charge can become zero after a partial conformal transformation, and vice versa. That is, the spacetime can become electrically and magnetically charged  from a vacuum neutral space via such a transformation. In this sense, whether or not one can experience electric or magnetic charge depends on the partial conformal factors.


  Based on the above discussions of the effects of partial conformal transformation for an Abelian gauge field, we further explore the corresponding effects for a
  non-Abelian field. We find a solution in the Yang-Mills-conformal gravity theory. Turning off the electromagnetic field in (\ref{lag}), we present a potential
  \be
  A=a_0\sigma_1 d\theta+(b_0\sin\theta \sigma_2+c_0\cos\theta\sigma_3)d\phi,
  \label{su2p}
  \en
  where $a_0,~b_0,~c_0$ are three constants, and $\sigma_i$ are the generators of the SU(2) group. A simple choice is the Pauli matrices. The field strength of the Yang-Mills reads,
  \bea
  F=dA+g_s A\wedge A= \left[(b_0-a_0c_0g_s)\cos\theta \sigma_2-(c_0-a_0b_0g_s)\sin\theta \sigma_3\right]d\theta\wedge d\phi.
  \label{fiel}
  \ena
  When $g_s=1$ and $a_0=b_0$, this solution reduces to a solution in \cite{lvhong} ($\psi=$constant and $g_s=1$ in their symbols). And in flat space, it is further degenerated to the Wu-Yang monopole \cite{wuyang}.
   It is easy to check that (\ref{su2p}) and (\ref{fiel}) satisfies the Yang-Mills equation, since
   \be
   dF=0,
   \en
   and
   \be
   A\wedge F=0.
   \en
  The electric charge
  \be
  q=\frac{1}{4\pi}\int (*F^A*F^A)^{1/2}=0,
  \en
  and the magnetic charge,
  \be
  U=\frac{1}{4\pi}\int (F^AF^A)^{1/2}=PE(1-\frac{Q^2}{P^2})+QE(1-\frac{P^2}{Q^2}),
  \label{YMcharge}
  \en
  where $E(x)$ presents the complete elliptic integral, and
  \be
  P=c_0-a_0b_0g_s,~Q=b_0-a_0c_0g_s.
  \en
  When $c_0=a_0b_0g_s$ and $b_0=a_0c_0g_s$ the Yang-Mills charge and the field strength vanish, and $A$ reduces to a pure gauge.
  We take the same metric as the case of an electromagnetic field, as shown in (\ref{schl}) and (\ref{fhi}).
  Then, mimicking the discussions in the case of an electromagnetic field, the field equation (\ref{bach}) requires,
  \be
  P^2=\frac{2\alpha(i^2-1-3hj)}{3 }.
  \label{cu}
  \en
  Thus we complete a solution in the Yang-Mills-conformal gravity. A special note is that we still do not obtain an exact solution in Einstein-Yang-Mills theory, for some related discussions see \cite{EYM, lvhong}. Our solution
  is not a solution of the  Einstein-Yang-Mills system.

  Comparing (\ref{cu}) with (\ref{qv}), one finds that the status of the magnetic charge of the Yang-Mills field is equal to the magnetic charge of the Maxwell field in the conformal gravity. After partial conformal transformations as in (\ref{tmet}) and (\ref{trans1}), the field equation (\ref{bach}) leads to,
  \be
  \widetilde{~P~}^2=-\frac{2\alpha \mu_3(\mu_1^2- \mu_2^2 i^2+3 \mu_2^2 h j)}{3 \mu_1^2\mu_2  }.
  \label{cup}
  \en
  The magnetic charge $U\to \widetilde{~U~}=\widetilde{~P~}E(1-\frac{Q^2}{P^2})+QE(1-\frac{P^2}{Q^2})$, which is similar to the case of the electromagnetic field. But there exists a difference. From (\ref{cu}), (\ref{YMcharge}) and (\ref{cup}) one may think that a partial  conformal transformation can make $P$ zero, but cannot completely obtain a zero Yang-Mills charge $U$.  Actually, $\widetilde{~U~}$ vanishes if one just sets $\widetilde{~P~}=-Q$. That is, to select a triplet ($\mu_1,~\mu_2,~\mu_3$) which satisfies
   \be
  \widetilde{~P~}=-Q=-\left(\left|\frac{2\alpha \mu_3(\mu_1^2- \mu_2^2 i^2+3 \mu_2^2 h j)}{3 \mu_1^2\mu_2  }\right|\right)^{1/2},
  \label{cup1}
  \en
  can lead to a zero Yang-Mills charge. The detailed discussions are similar to the case of the electromagnetic field. So, whether or not there is a SU(2) Yang-Mills field depends on the partial conformal factors.


  In summary, we find a new property of the conformal gravity. We demonstrate that the charge, and thus the gauge field itself, depends on the partial conformal factors. In a Maxwell-Yang-Mills-conformal gravity system, the conformal symmetry is preserved if the gauge fields involved in are source-free. For such a system, a (total) conformal transformation yields no physical effects. We introduce the concept ``partial conformal transformation", that is, the transformations in subspaces of the spacetime. Through studies in the spherically symmetric spacetimes, we find that both the U(1) and SU(2) charges are dependent on the selection of the partial conformal factors.

      The static solution of the Einstein equation with U(1) gauge field, i.e., the Reissner-Nordstr$\ddot{o}$m solution \cite{RN} was obtained 100 years ago. The SU(n)-colored black holes also have been investigated for several years \cite{EYM}. But in all previous cases, the gauge fields must be introduced by hand. In comparison, in our progress they spontaneously appear after a partial conformal symmetry from a neutral solution. For example, if $i^2-1-3hj=0$, then (\ref{qv}) indicates that the solution is a neutral vacuum solution. After  partial conformal transformations, the electromagnetic charges appear. And the extra stress energy just satisfies the field equation with the metric after transformations, and the electromagnetic field just satisfies the Maxwell equation. Therefore, one could say that whether one can sense an electromagnetic field depends on the selection of partial conformal factors. Similarly, for the SU(2)-colored black hole (\ref{su2p}), we obtain different charges if we adopt different rescale factors $\mu_1,~\mu_2,~\mu_3$. Based on the past experiences, a varied quantity under some transformation may be a component of an invariant or covariant quantity. To find such a covariant quantity is an interesting topic.

   The conformal gravity takes a special status in gravity theories. The conformal symmetry is shared by all standard model particles at enough high energy and thus may be a property of the unification theory including gravity and gauge interactions.  In this sense, we can say that the conformal gravity may encode some essences of the envisaged unification theory. Thus, we further conjecture that the gauge field theory may be an effect of some scaling symmetry of spacetime in the unification theory including gauge and gravity theories. A  note is that the subspaces involved in the transformations in this paper make physical senses. $\frac{\partial}{\partial t}$ is the unique time-like Killing field, and $r,~\theta,~\phi$ are the area coordinates of the spacetime. The relation between gauge fields and partial conformal transformations in different subspaces needs to be explored further.

   It deserves to recapitulate the partial conformal transformation in a wider background. Because we adopt metric-compatible derivative operators both before and after partial transformations, the derivative operator is uniquely determined by the metric. Considering the Weyl's original suggestion, it is natural to extend our discussions to metric-affine gravity. In metric-affine gravity, the metric and derivative operator (connection) are defined independently. Thus a metric cannot uniquely determine a connection \cite{rev1}. For the vacuum Einstein-Hilbert action, a variation with respect to the connection leads to  standard Christoffel symbols.   The presence of matters changes the situation, and the field equation gets essentially modified. The independent connections (torsion is permitted) introduce 64 new degrees in 4 dimensional spacetime, which leads to considerable additional flexibility and at the same time, uncertainty. An interesting relation between metric-affine gravity and partial conformal transformation deserves to note is that partial conformal transformations can naturally lead to a (special) metric-affine theory.  This point can be clearly seen from the following arguments \cite{earsum}. The group of partial conformal transformations is a subgroup of the general linear transformation group GL(4,R). When the general linear group GL(4,R) is localized and gauged, one arrives at a metric-affine gravity. Thus, when the group of the partial conformal transformations is localized and gauged, one reaches to a metric-affine gravity, which is a special case of the former one. To perform partial transformations in metric-affine gravity, one needs a concrete solution. One usually needs to introduce more extra constraints by hand to construct a concrete solution concerning this uncertainty of metric-affine gravity. Some early solutions, including dilation, monopole, spin-charged, and cosmological ones, etc,  in metric-affine gravity is summarized in \cite{earsum}. Geonic wormholes in metric-affine gravity are thoroughly investigated in \cite{drthesis}. Under the ``triplet ansatz", some static black holes in presence of Yang-Mills matters are obtained in \cite{ABea}, and the related junction condition is studied in \cite{junction}. To perform partial conformal transformations to these solutions is an interesting topic and a natural extension of the present study, which is expected to have much rich structures.  We shall carefully investigate this point in the coming work.

  We make a preliminary unification of the electromagnetic, SU(2), and conformal gravity. How far can we reach when we extend this argument to the Einstein gravity? First we write the Ricci scalar before the partial conformal transformations for the general spherical metric (\ref{schl}),
  \be
  R=\frac{1}{r^2}(2-2f-4rf'-r^2f'').
  \en
  After partial conformal transformations, by using (\ref{tmet}) the Ricci scalar is shifted to be,
  \be
  R=\frac{1}{\mu_1\mu_2 r^2}(2\mu_1-2\mu_2f-4\mu_2rf'-\mu_2r^2f'').
  \en
 Thus, the Einstein-Hilbert action is not invariant unless $\mu_1=\mu_2$. From the discussions on the conformal gravity, we see that this shift does not necessarily imply that the Einstein gravity cannot realize the similar property in conformal gravity. Now we check the transformation properties of the field equation and the stress tensor for the Reissner-Nordstr$\ddot{o}$m metric with magnetic charges. The metric reads,
 \be
 f=1-\frac{2m}{r}+\frac{q^2+v^2}{r^2},
 \en
 and the source reads,
 \be
 {\cal A}=\frac{q}{r}dt-v\cos\theta d\phi,
 \en
 which is the same as the case of the conformal gravity. After the partial conformal transformations, the metric becomes (\ref{tmet}). Based on this metric, one obtains,
 \be
 G^0_0-8\pi T^0_0=\frac{(\mu_2\mu_3-\mu_0\mu_2\mu_3)q^2+(\mu_0\mu_1-\mu_0\mu_2\mu_3)v^2+(\mu_0\mu_2\mu_3-\mu_0\mu_1\mu_3)r^2}{\mu_0\mu_1\mu_2\mu_3 r^4},
 \en
 \be
 G^1_1-8\pi T^1_1=G^0_0-8\pi T^0_0,
 \en
 \be
 G^2_2-8\pi T^2_2=\frac{-(\mu_2\mu_3-\mu_0\mu_2\mu_3)q^2-(\mu_0\mu_1-\mu_0\mu_2\mu_3)v^2}{\mu_0\mu_1\mu_2\mu_3 r^4},
 \en
 and,
 \be
 G^3_3-8\pi T^3_3=G^2_2-8\pi T^2_2.
 \en
 One sees that a parameter array ($\mu_0,~\mu_1,~\mu_2,~\mu_3$) can vanish $G_2^2-8\pi T_2^2$ and $G_3^3-8\pi T_3^3$, but never does so for $G_0^0-8\pi T_0^0$ and $G_1^1-8\pi T_1^1$, since the existence of the $r^2$-term in the numerator. Thus, generally, the metric after partial conformal transformations can no longer solve the field equation whatever how to choose the parameters ($\mu_0,~\mu_1,~\mu_2,~\mu_3$). This is different from the case of conformal gravity. We expect that partial conformal transformations with more freedoms may do this work. For the Yang-Mills colored black hole in Einstein gravity, we still do not have an exact solution. It seems difficult to perform similar transformations for a numerical solution.

   Now we make a final discussion based on some recent progresses of t' Hooft. The conformal gravity emerges naturally when one considers the renormalization of the matter fields. When one consider quantum effects of the matter fields, a non-minimally coupling term to the gravity field appears \cite{BD}. And the conformal gravity is dictated by the renormalization of a non-minimally coupled matter field \cite{hoo1, hoo2}. Both the matter fields and gravity field may have conformal symmetries in some high energy region. In the low energy region the conformal symmetry is spontaneously broken and the Einstein gravity recovers \cite{hoo}. Thus the unification discussed in this paper may be achieved in some high energy scale, but is broken in the present energy scale.

 {\bf Acknowledgments.}
   We thank the anonymous reviewer for her/his several valuable comments and questions. This work is supported in part by the National Natural Science Foundation of China (NSFC) under grant No. 11575083, and  Shandong Province Natural Science Foundation under grant No.  ZR201709220395.

\end{document}